\documentclass[prb,twocolumn,superscriptaddress,floatfix,showkeys]{revtex4}
\usepackage{graphicx}
\usepackage{amsmath,subfigure,epsfig,psfrag}
\usepackage{comment}
\usepackage{commath}
\usepackage{graphicx}
\usepackage{amssymb}
\usepackage{xcolor}
\usepackage{soul}
\usepackage{listings}
\usepackage{rotating}
\usepackage{latexsym}
\usepackage{amsfonts}
\usepackage{amssymb}
\usepackage{wasysym}
\usepackage{xfrac}
\usepackage{mathtools}
\usepackage{multirow}
\makeatletter
\renewcommand*\env@matrix[1][\arraystretch]{%
  \edef\arraystretch{#1}%
  \hskip -\arraycolsep
  \let\@ifnextchar\new@ifnextchar
  \array{*\c@MaxMatrixCols c}}
\makeatother

\begin{document}

\title{Structural Relaxation of Materials with Spin-Orbit Coupling: Analytical Forces in Spin-Current DFT}

\author{Jacques K. Desmarais}
\email{jacqueskontak.desmarais@unito.it}
\affiliation{Dipartimento di Chimica, Universit\`{a} di Torino, via Giuria 5, 10125 Torino, Italy}

\author{Alessandro Erba}
\email{alessandro.erba@unito.it}
\affiliation{Dipartimento di Chimica, Universit\`{a} di Torino, via Giuria 5, 10125 Torino, Italy}

\author{Jean-Pierre Flament}
\affiliation{Universit{\'e} de Lille, CNRS, UMR 8523 --- PhLAM --- Physique des Lasers, Atomes et Mol{\'e}cules, 59000 Lille, France}

\date{\today}

\begin{abstract}
Analytical gradients of the total energy are provided for local density and generalized-gradient hybrid approximations to generalized Kohn-Sham spin-current density functional theory (SCDFT). It is shown that gradients may be determined analytically, in a two-component framework, including spin-orbit coupling (SOC), with high accuracy. We demonstrate that renormalization of the electron-electron potential by SOC-induced spin-currents can account for considerable modification of crystal structures. In the case of Iodine-based molecular crystals, the effect may amount to more than half of the total modification of the structure by SOC. Such effects necessitate an SCDFT, rather than DFT, formulation, in which exchange-correlation functionals are endowed with an explicit dependence on spin-current densities. An implementation is presented in the \textsc{Crystal} program.
\end{abstract}

\maketitle

\section{Introduction}

The Hohenberg-Kohn density functional theory (DFT), being entirely formulated in terms of functionals $F \left[ n \right]$ of the electron density (particle density) $n=\Psi^\dagger \Psi$, is meant for a (non-relativistic) fermionic system embedded in some external field which may be described by a scalar-multiplicative potential (i.e. a Coulomb field).\cite{hohenberg1964inhomogeneous} The theory may be extended to external fields that are not scalar-multiplicative, a consequence being that the formulation, then, involves a larger set of auxiliary density variables. For instance, extension to a Zeeman field leads to functionals $F \left[ n,\mathbf{m} \right]$ of both the electron density and spin-magnetization $\mathbf{m}=\Psi^\dagger \boldsymbol{\sigma} \Psi$, where $\boldsymbol{\sigma}$ is the vector of Pauli matrices $\boldsymbol{\sigma}^x, \boldsymbol{\sigma}^y$ and $\boldsymbol{\sigma}^z$: i.e. the so-called spin-DFT (SDFT) of von Barth and Hedin.\cite{von1972local} Further extension to an external magnetic field leads to the current-spin DFT of Vignale and Rasolt,\cite{vignale1987density,vignale1988current} involving functionals  $F \left[ n,\mathbf{m},\mathbf{j} \right]$ of also the particle-current density $\mathbf{j}= \frac{1}{2i} \Psi^\dagger \left( \boldsymbol{\nabla} -  \boldsymbol{\nabla}^\dagger \right) \Psi$. The appearance of $\mathbf{m}$ and $\mathbf{j}$ in the formulation is thus associated to time-reversal symmetry breaking (TRSB) due to the magnetic field. In (open-shell) systems that intrinsically break TRS, use of SDFT has become routine.

The above considerations are, of course, still restricted to the non-relativistic regime. Scalar relativistic (SR) effects (by definition described by scalar-multiplicative potentials) can be straightforwardly included in the theory. It is notable, however, that spin-orbit coupling (SOC, described by a potential that is certainly not scalar,  nor multiplicative) may also be formally introduced in the theory by viewing it as another external field.  Overall, the relativistic generalization of the procedure leads (in the two-component framework) to the spin-current DFT (SCDFT), of Vignale and Rasolt, as first shown by Bencheihk.\cite{vignale1988current,bencheikh2003spin} In this context, SOC enters the theory through non-Abelian potentials $\boldsymbol{\mathcal{A}}^x$, $\boldsymbol{\mathcal{A}}^y$ and $\boldsymbol{\mathcal{A}}^z$, each of which couple to the fermionic system through spin-current densities (i.e. currents of the spin-magnetization $m_x$, $m_y$ and $m_z$)  $\mathbf{J}^x$, $\mathbf{J}^y$ and $\mathbf{J}^z$:\cite{bencheikh2003spin,pittalis2017u,desmarais2020adiabatic,desmarais2020spin} 
\begin{equation}
\mathbf{J}^a=\frac{1}{2i} \Psi^\dagger \boldsymbol{\sigma}^a  \left( \boldsymbol{\nabla} -  \boldsymbol{\nabla}^\dagger \right) \Psi \quad a=x,y,z
\end{equation}
In the simplest variant of the formulation, that being for (closed-shell) systems that preserve time-reversal symmetry, the functional reduces to $F \left[ n,\mathbf{J}^x,\mathbf{J}^y,\mathbf{J}^z \right]$, and therefore corresponding density-functional approximations (DFAs) for a fermionic system in the presence of SOC must include the spin-current densities.\cite{trushin2018spin,bodo2022spin} In the most general case of TRSB systems, the full functional  $F \left[ n,\mathbf{m},\mathbf{j},\mathbf{J}^x,\mathbf{J}^y,\mathbf{J}^z \right]$ is written also in terms of $\mathbf{m}$ and $\mathbf{j}$.

Although first proposed around 35 years ago, SCDFT has garnered little attention, until recently. Indeed, the ball was moved forward in 2017, with Pittalis {\it et al.} demonstrating that the spin-currents enter explicitly in DFAs only at the level of the curvature of the exchange-correlation (xc) hole.\cite{pittalis2017u} This led to our formulation of the adiabatic connection in the SCDFT, within a generalized Kohn-Sham framework,\cite{seidl1996generalized} in which spin-currents can be effectively treated via the exact-exchange operator in hybrid DFAs of the local-density and generalized gradient approximations (LDA and GGA).\cite{desmarais2020adiabatic,desmarais2020spin} The theory was thereafter applied to the description of: i) Weyl fermions, wherein renormalization of the electron-electron interaction by SOC-induced spin-currents was found to account for around half of the splitting of the Weyl node pair in TaAs;\cite{bodo2022spin} ii) a Bismuth two-dimensional $\mathbb{Z}_2$ topological insulator, wherein it was demonstrated that only an SCDFT formulation could account for the appearance of an experimentally-confirmed Dirac fermion in the valence band structure, at the onset of the topological phase transition.\cite{comaskey2022spin}

These previous studies clearly demonstrate the fundamental importance of spin-current densities in the description of the electronic structure, when SOC is included. 
Here, we extend the same treatment to analytical gradients of the total energy, allowing, for the first time, to discuss the effect of spin-currents on the description of the crystal structure. Such extension is implemented in a developmental version of the \textsc{Crystal} program.\cite{CRYSTAL17PAP,erba2022crystal23}  Through application of the approach to the diiodide molecule, as well as Iodine-based molecular crystals, we demonstrate that renormalization of the electron-electron interaction through SOC-induced spin-currents can account for significant modification of crystal structures (around half or more of the total effect due to SOC).

\section{Formalism}
\label{sec:form}

\subsection{Two-Component Generalized Kohn-Sham Equations}

In the case of periodic systems, the spinors $\vert \psi_{i,\mathbf{k}} \rangle = \vert \psi_{i,\mathbf{k}}^\uparrow \rangle \otimes \vert \uparrow \rangle + \vert \psi_{i,\mathbf{k}}^\downarrow \rangle \otimes \vert \downarrow \rangle$  are 2c crystalline orbitals (COs), with components $\vert \psi_{i,\mathbf{k}}^\sigma \rangle$, expanded in a set of Bloch functions (BFs) $\phi_{\mu,\mathbf{k}}$:
\begin{equation}
\label{eqn:co}
\psi_{i,\mathbf{k}}^\sigma \left( \mathbf{r} \right) = \sum_\mu^{N_\mathcal{B}} C^{\sigma}_{\mu,i} \left( \mathbf{k} \right) \phi_{\mu,\mathbf{k}} \left( \mathbf{r} \right) \; ,
\end{equation}
where $\mathbf{k}$ is a point in the first Brillouin zone (FBZ), $N_\mathcal{B}$ is the number of basis functions in a given cell of the infinite-periodic system, and $\sigma=\uparrow, \downarrow$ is a spin index.

In \textsc{Crystal}, the BFs are conveniently represented as a linear combination of \textit{pure real} atomic orbitals (LCAO), through the inverse-Fourier relation:
\begin{equation}
\label{eqn:Bloch}
\phi_{\mu,\mathbf{k}} \left( \mathbf{r} \right) = \frac{1}{\sqrt{\Omega}} \sum_{\mathbf{g}} e^{\imath \mathbf{k} \cdot \mathbf{g}} \ \chi_{\mu,\mathbf{g}} \left( \mathbf{r} - \mathbf{A}_\mu \right) \; .
\end{equation}
Here $\Omega$ is the volume of the FBZ, $\mathbf{g}$ is a direct-lattice vector and $\mathbf{A}_\mu$ is the position in cell $\mathbf{g}$ at which the AO $\chi_{\mu,\mathbf{g}}$ is centered. In Eq. (\ref{eqn:Bloch}) we have introduced the shorthand notation $\chi_{\mu,\mathbf{g}} \left( \mathbf{r} - \mathbf{A}_\mu \right)=\chi_{\mu} \left( \mathbf{r} - \mathbf{A}_\mu - \mathbf{g} \right)$. A similar notation will also be applied to the electron density $\rho_{\mathbf{g}} \left( \mathbf{r} \right)=\rho \left( \mathbf{r}-\mathbf{g} \right)$ Variation of the orbitals $\psi_{i,\mathbf{k}}^\sigma$ under the constraint of orthonormality:
\begin{subequations}
\begin{eqnarray}
\label{eqn:ortho}
\langle \psi_{i,\mathbf{k}}^\sigma \vert \psi_{j,\mathbf{k}^\prime}^{\sigma^\prime} \rangle &=& \delta_{i,j} \delta_{\mathbf{k},\mathbf{k}^\prime} \delta_{\sigma,\sigma^\prime} \nonumber \\
&\Rightarrow& \mathbf{C}^\dagger \left( \mathbf{k} \right) \mathbf{S} \left( \mathbf{k} \right) \mathbf{C} \left( \mathbf{k} \right) = \mathbf{1}
\end{eqnarray}
leads to the generalized Kohn-Sham (GKS) equation:
\begin{equation}
\label{eqn:GKS}
\mathbf{H} \left( \mathbf{k} \right)  \mathbf{C} \left( \mathbf{k} \right)  = \mathbf{S} \left( \mathbf{k} \right)  \mathbf{C} \left( \mathbf{k} \right)  \mathbf{E} \left( \mathbf{k} \right) \; ,
\end{equation}
\end{subequations}
where all matrices have size $2N_\mathcal{B} \times 2N_\mathcal{B}$, $\mathbf{C} \left( \mathbf{k} \right)$ is the matrix of CO coefficients of Eq. (\ref{eqn:co}), $\mathbf{S} \left( \mathbf{k} \right)$ is the BF overlap matrix, $\mathbf{E} \left( \mathbf{k} \right)$ is the matrix of Lagrange multipliers (i.e. for canonical orbitals, corresponding to the diagonal matrix of band-structure energy levels $\epsilon_{i,\mathbf{k}}$) and $\mathbf{H} \left( \mathbf{k} \right)$ is the BF Hamiltonian matrix. Eq. (\ref{eqn:GKS}) can be written more explicitly to highlight the structure in spin space:
\begin{eqnarray}
\label{eqn:GKS_explicit}
&&
\begin{pmatrix}
\mathbf{H}^{\uparrow \uparrow} \left( \mathbf{k} \right) & \mathbf{H}^{\uparrow \downarrow} \left( \mathbf{k} \right)  \\
\mathbf{H}^{\downarrow \uparrow} \left( \mathbf{k} \right) & \mathbf{H}^{\downarrow \downarrow} \left( \mathbf{k} \right)
\end{pmatrix}
\begin{pmatrix}
\mathbf{C}^{\uparrow} \left( \mathbf{k} \right) \\
\mathbf{C}^{\downarrow} \left( \mathbf{k} \right)
\end{pmatrix}
\nonumber \\
&=& 
\begin{pmatrix}
\mathbf{S}^{\uparrow \uparrow} \left( \mathbf{k} \right) & \mathbf{0} \\
\mathbf{0} & \mathbf{S}^{\downarrow \downarrow} \left( \mathbf{k} \right)
\end{pmatrix}
\begin{pmatrix}
\mathbf{C}^{\uparrow} \left( \mathbf{k} \right) \\
\mathbf{C}^{\downarrow} \left( \mathbf{k} \right)
\end{pmatrix}
\mathbf{E} \left( \mathbf{k} \right) \; .
\end{eqnarray}
In Eq. (\ref{eqn:GKS_explicit}) and elsewhere, matrices with double and single spin indices have size $N_\mathcal{B} \times N_\mathcal{B}$ and $N_\mathcal{B} \times 2N_\mathcal{B}$, respectively. $\mathbf{H}^{\sigma \sigma^\prime} \left( \mathbf{k} \right)$, for instance, has elements:
\begin{equation}
\label{eqn:fock_BF}
H^{\sigma \sigma^\prime}_{\mu \nu} \left( \mathbf{k} \right)= \Omega \langle \phi_{\mu,\mathbf{k}} \vert \hat{H}^{\sigma \sigma^\prime} \vert \phi_{\nu,\mathbf{k}} \rangle
\end{equation}
and:
\begin{equation}
\label{eqn:fock}
\mathbf{H}^{\sigma \sigma^\prime} \left( \mathbf{k} \right) = \mathbf{h}^{\sigma \sigma^\prime} \left( \mathbf{k} \right) + \mathbf{J}^{\sigma \sigma^\prime} \left( \mathbf{k} \right)- a \mathbf{K}^{\sigma \sigma^\prime} \left( \mathbf{k} \right)+ \mathbf{V}^{\sigma \sigma^\prime} \left( \mathbf{k} \right)\;,
\end{equation}
in which $\mathbf{h}^{\sigma \sigma^\prime} \left( \mathbf{k} \right)$ contains the matrix elements that can be built from mono-electronic integrals:
\begin{equation}
\label{eqn:fock_1e}
\mathbf{h}^{\sigma \sigma^\prime} \left( \mathbf{k} \right) =\delta_{\sigma,\sigma^\prime} \left[ \mathbf{v} \left( \mathbf{k} \right)+ \mathbf{u}_{AR} \left( \mathbf{k} \right)\right]+\mathbf{u}_{SO}^{\sigma \sigma^\prime} \left( \mathbf{k} \right)\;.
\end{equation}
Here, $\mathbf{v}$ consists of the electronic kinetic energy and electron-nuclear interaction terms, $\mathbf{u}_{AR}$ and $\mathbf{u}_{SO}^{\sigma \sigma^\prime}$ are, respectively, the averaged and spin-orbit relativistic effective potential (AREP and SOREP) matrices;\cite{ermler1981ab,cao2011pseudopotentials} and $\mathbf{J}^{\sigma \sigma^\prime}$ and $\mathbf{K}^{\sigma \sigma^\prime}$ are the usual Coulomb and exact-exchange terms (with $a$ being the included fraction of the latter). $\mathbf{V}^{\sigma \sigma^\prime}$ is the matrix of DFT correlation and exchange potentials (in either collinear or non-collinear treatments).\cite{desmarais2021perturbationII,desmarais2021spin}

Inserting Eq. (\ref{eqn:Bloch}) into Eq. (\ref{eqn:fock_BF}) (or the equivalent equation with $\hat{H}$ being replaced by any other operator), we are able to relate the BF matrix $\mathbf{H}^{\sigma \sigma^\prime} \left( \mathbf{k} \right)$, for instance, to the AO one $\mathbf{H}^{\sigma \sigma^\prime} \left( \mathbf{g} \right)$ through the inverse-Fourier relation:
\begin{subequations}
\begin{equation}
\label{eqn:fourier}
\mathbf{H}^{\sigma \sigma^\prime} \left( \mathbf{k} \right) = \sum_{\mathbf{g}} e^{\imath \mathbf{k} \cdot \mathbf{g}} \ \mathbf{H}^{\sigma \sigma^\prime} \left( \mathbf{g} \right) \; ,
\end{equation}
where AO matrix elements of $\hat{H}$ or any other operator read:
\begin{equation}
\label{eqn:Hg}
\mathbf{H}^{\sigma \sigma^\prime} \left( \mathbf{g} \right) = \langle \chi_{\mu,\mathbf{0}} \vert \hat{H}^{\sigma \sigma^\prime} \vert \chi_{\nu,\mathbf{g}} \rangle \; .
\end{equation}
\end{subequations}
In the AO basis, the Coulomb matrix reads:\cite{erba2022crystal23}
\small
\begin{eqnarray}
\label{eqn:coul}
J^{\sigma \sigma^\prime}_{\mu \nu} \left( \mathbf{g} \right) &=& \delta_{\sigma,\sigma^\prime} \sum_{\tau \omega} \sum_\mathbf{g'} \Re \left[ P_{\omega \tau}^{\uparrow \uparrow} \left( \mathbf{g'} \right)  + P_{\omega \tau}^{\downarrow \downarrow} \left( \mathbf{g'} \right)  \right] \sum_{\mathbf{g''}} (\mu^{\mathbf{0}} \nu^{\mathbf{g}} |\tau^{\mathbf{g''}} \omega^{\mathbf{g''}+\mathbf{g'}} ) \nonumber \\
&=&  \delta_{\sigma,\sigma^\prime}  \sum_{\mathbf{g''}} \int d\mathbf{r} \ \chi_{\mu,\mathbf{0}} \left( \mathbf{r} - \mathbf{A}_\mu \right) \Phi^\text{Coul} \left( \mathbf{r},\mathbf{g''} \right) \chi_{\nu,\mathbf{g}} \left( \mathbf{r} - \mathbf{A}_\nu \right) \; ,
\end{eqnarray}
\normalsize
with the Coulomb potential:
\begin{equation}
\label{eqn:coul_pot}
\Phi^\text{Coul} \left( \mathbf{r},\mathbf{g''} \right) = \int d\mathbf{r}^\prime \frac{\rho_{\mathbf{g''}} \left( \mathbf{r}^\prime\right)}{ \vert \mathbf{r}^\prime - \mathbf{g''} - \mathbf{r} \vert}
\end{equation}
being a density functional $\Phi^\text{Coul} = \Phi^\text{Coul}\left[ \rho_\mathbf{g''}  \right]$. The exchange AO matrix reads:
\begin{equation}
\label{eqn:exch}
K^{\sigma \sigma^\prime}_{\mu \nu} \left( \mathbf{g} \right)=\sum_{\tau \omega} \sum_\mathbf{g'} P_{\tau \omega}^{\sigma \sigma^\prime} \left( \mathbf{g'} \right) \sum_{\mathbf{g''}} (\mu^{\mathbf{0}} \tau^{\mathbf{g''}} |\omega^{\mathbf{g''}+\mathbf{g'}} \nu^\mathbf{g}) \; ,
\end{equation}
where $P_{\mu \nu}^{\sigma \sigma^\prime} \left( \mathbf{g} \right)$ is the AO direct-space density matrix:
\begin{eqnarray}
\label{eqn:pmat}
P_{\mu \nu}^{\sigma \sigma^\prime} \left( \mathbf{g} \right) &=& \left[ P_{\nu \mu}^{\sigma^\prime \sigma} \left( -\mathbf{g} \right)\right]^\ast = \frac{1}{\Omega} \sum_i^\text{bands} f_i \int_\Omega d \mathbf{k} \ e^{\i \mathbf{k} \cdot \mathbf{g}} \nonumber \\
&\times&C^{\sigma}_{\mu,i} \left( \mathbf{k} \right) \left[ C^{\sigma^\prime}_{\nu,i} \left( \mathbf{k} \right) \right]^\ast \theta\left[ \varepsilon_F - \epsilon_i ({\mathbf{k}}) \right]  \; ,
\end{eqnarray}
and $\theta$ is the Heaviside step-function, $\varepsilon_F$ is the Fermi energy and $0 < f_{i} < 1$ is the fractional occupation of band $i$. In terms of these matrices, the total energy is written:
\begin{eqnarray}
\label{eqn:E}
E &=& \frac{1}{2} \sum_\mathbf{g}   \sum_{\sigma,\sigma^\prime} \sum_{\mu \nu}   \Bigg\{ \left[ P_{\mu \nu}^{\sigma \sigma^\prime} \left( \mathbf{g} \right) \right]^\ast  \left[  h^{\sigma \sigma^\prime}_{\mu \nu} \left( \mathbf{g} \right)  + H^{\sigma \sigma^\prime}_{\mu \nu} \left( \mathbf{g} \right) \right] \Bigg\} \nonumber \\
&=& \frac{1}{2} \sum_\mathbf{g}  \sum_{\sigma,\sigma^\prime} \sum_{\mu \nu}  \Bigg\{ \left[   P_{\mu \nu}^{\sigma \sigma^\prime} \left( \mathbf{g} \right) \right]^\ast  \Bigg[ 2 h^{\sigma \sigma^\prime}_{\mu \nu} \left( \mathbf{g} \right) + V^{\sigma \sigma^\prime}_{\mu \nu} \left( \mathbf{g} \right) \nonumber \\
&+& \sum_{\tau \omega} \sum_{\sigma^{\prime \prime},\sigma^{\prime \prime \prime}}  \sum_\mathbf{g'} P_{\tau \omega}^{\sigma^{\prime \prime} \sigma^{\prime \prime \prime}} \left( \mathbf{g'} \right) \sum_{\mathbf{g''}} B_{\mu,\nu,\tau,\omega}^\mathbf{0,g,g'',g''+g'} \Bigg] \Bigg\} \; ,
\end{eqnarray} 
where we introduced the shorthand notation:
\begin{equation}
B_{\mu,\nu,\tau,\omega}^\mathbf{0,g,g'',g''+g'}= (\mu^{\mathbf{0}} \nu^{\mathbf{g}} | \tau^{\mathbf{g''}} \omega^{\mathbf{g''}+\mathbf{g'}} )  -a (\mu^{\mathbf{0}} \tau^{\mathbf{g''}} | \omega^{\mathbf{g''}+\mathbf{g'}} \nu^\mathbf{g}) \; .
\end{equation}

\subsection{Treatment of the Coulomb Series}

For 3D periodic systems, the Coulomb lattice series, whose electron-electron component was given in Eq. (\ref{eqn:coul}), is conditionally convergent. The series may be rendered absolutely convergent by Ewald summation techniques, employing a charge distribution of atom-centered point multipoles (here $c$ is an atomic index per cell) in the long range:\cite{saunders1992electrostatic,CRYSTAL88_1}
\begin{equation}
\label{eqn:rho_model}
\rho^\text{model}_{c,\mathbf{g''}} \left( \mathbf{r} \right) = \sum_{l=0}^L \sum_{m=-l}^l \eta_m^l \left[ \rho_{c,\mathbf{g''}} \right] \delta_l^m \left( \mathbf{r}-\mathbf{A}_c-\mathbf{g''} \right) \; ,
\end{equation}
used to model the exact atomic contribution to the density:
\begin{eqnarray}
\label{eqn:rho_c}
\rho_{c,\mathbf{g''}} \left( \mathbf{r} \right) &=& \sum_{\mu \in {c,\mathbf{g''}}} \sum_\mathbf{g} \sum_\nu \Re \left[ P_{\mu \nu}^{\uparrow \uparrow} \left( \mathbf{g} \right)  + P_{\mu \nu}^{\downarrow \downarrow} \left( \mathbf{g} \right)  \right] \nonumber \\
&\times& \chi_\mu \left( \mathbf{r} -\mathbf{A}_c-{\mathbf{g''}} \right) \chi_\nu \left( \mathbf{r} -\mathbf{A}_\nu - \mathbf{g}-\mathbf{g''} \right) \; ,
\end{eqnarray}
with $\mu \in {c,\mathbf{g''}}$ meaning that the sum is restricted to AOs centered at atom $c$ in cell $\mathbf{g''}$. In Eq. (\ref{eqn:rho_model}):
\begin{equation}
\eta_m^l \left[ \rho_{c,\mathbf{g''}} \right] = \int d \mathbf{r} \rho_{c,\mathbf{g''}} \left( \mathbf{r} \right) X_l^m \left( \mathbf{r} - \mathbf{A}_c-\mathbf{g''} \right)
\end{equation}
are the multipole moments of $\rho_{c,\mathbf{g''}}$ with unnormalized real spherical harmonics $X_l^m$, while $\delta_l^m$ are unit point multipoles centered at $\mathbf{A}_c$ in cell $\mathbf{g''}$, and in our implementation $L$ has a maximum value of 6 (a minimum value of 4 is formally required to ensure absolute convergence).

The model density is introduced by the following replacement of the Coulomb potential in Eq. (\ref{eqn:coul}):\cite{saunders1992electrostatic}
\begin{equation}
\label{eqn:coul_ew}
\Phi^\text{Coul} \left[ \rho_{c,\mathbf{g''}} \right] \to \Phi^\text{Ew} \left[ \rho^\text{model}_{c,\mathbf{g''}} \right] + \Phi^\text{Coul} \left[ \rho_{c,{\mathbf{g''}}} - \rho^\text{model}_{c,{\mathbf{g''}}} \right] \; ,
\end{equation}
with $\Phi^\text{Ew}$ being the corresponding Ewald potential. The model $\rho^\text{model}_{c,\mathbf{g''}}$ is applied in the long range, in the sense that the ${\mathbf{g''}}$ lattice series for the second term in Eq. (\ref{eqn:coul_ew}) is truncated by a preset tolerance $T2$, while the one relevant to the first term is summed analytically to infinity.\cite{saunders1992electrostatic,CRYSTAL88_1}

In the 3D periodic case, the procedure leads to an absolutely convergent lattice series, at the price of an additional correction depending on the shape of the sample used for the summation.\cite{de1980simulation} For spherical 3D samples, the correction to the Hamiltonian matrix is proportional to the spherical second moment of the electron density $Q$:\cite{saunders1992electrostatic}
\begin{equation}
\label{eqn:H_mod}
H_{\mu \nu}^{\sigma \sigma^\prime} \left( \mathbf{g} \right) \to H_{\mu \nu}^{\sigma \sigma^\prime} \left( \mathbf{g} \right) - \delta_{\sigma, \sigma^\prime} Q S_{\mu \nu}^{\sigma \sigma} \left( \mathbf{g} \right) \; ,
\end{equation}
where:
\begin{equation}
\label{eqn:sphero}
Q = \sum_c^\text{atoms} Q_c = \sum_c^\text{atoms} \frac{2 \pi}{3 V} \int d \mathbf{r} \left[ \rho_{c,{\mathbf{0}}} \left( \mathbf{r} \right)  - \rho^\text{model}_{c,{\mathbf{0}}} \left( \mathbf{r} \right)\right] \vert \mathbf{r} \vert^2 \; ,
\end{equation}
and $V$ is the volume of the unit cell in direct space.

In the calculation, the correction of the Hamiltonian matrix elements of Eq. (\ref{eqn:H_mod}) can be avoided by adding $Q \mathbf{S} \left( \mathbf{k} \right)$ to both sides of Eq. (\ref{eqn:GKS}), leading to the modified GKS equation:\cite{saunders1992electrostatic}
\begin{equation}
\label{eqn:GKS_mod}
\mathbf{H} \left( \mathbf{k} \right) \mathbf{C} \left( \mathbf{k} \right)  = \mathbf{S} \left( \mathbf{k} \right)  \mathbf{C} \left( \mathbf{k} \right)  \left( \mathbf{E} \left( \mathbf{k} \right) + Q \right) \; ,
\end{equation}
with shifted energy levels $\mathbf{E} \left( \mathbf{k} \right) \to \mathbf{E} \left( \mathbf{k} \right) + Q$. The shift is irrelevant for total energy calculations (in which only differences $\epsilon_i \left( \mathbf{k} \right) - \varepsilon_F$ with respect to the Fermi level $\varepsilon_F$ matter). On the other hand, the correction enters our formulation for the analytical gradient in 3D periodic systems, as shown below. We note that the same correction is not necessary for the 0D, 1D or 2D cases, in which the Coulomb lattice series is already absolutely convergent (provided that a unit cell can be chosen with vanishing dipole moment).

\subsection{Analytical Gradients with Respect to Atomic Displacements}
\label{sec:grad_atom}

The derivative of the energy $E$ with respect to one of the atomic centers $\mathbf{A}_\eta$ provides, from Eq. (\ref{eqn:E}):
\begin{eqnarray}
\label{eqn:gradE}
\frac{\partial E}{\partial \mathbf{A}_\eta} &=& \frac{1}{2} \sum_\mathbf{g}  \sum_{\sigma,\sigma^\prime} \sum_{\mu \nu}  \Bigg\{  \left[ P_{\mu \nu}^{\sigma \sigma^\prime} \left( \mathbf{g} \right) \right]^\ast \Bigg[ 2 \frac{\partial  h^{\sigma \sigma^\prime}_{\mu \nu} \left( \mathbf{g} \right)  }{\partial \mathbf{A}_\eta} + \frac{\partial  V^{\sigma \sigma^\prime}_{\mu \nu} \left( \mathbf{g} \right)  }{\partial \mathbf{A}_\eta} \nonumber \\
&+& \sum_{\tau \omega} \sum_{\sigma^{\prime \prime},\sigma^{\prime \prime \prime}}  \sum_\mathbf{g'} P_{\tau \omega}^{\sigma^{\prime \prime} \sigma^{\prime \prime \prime}} \left( \mathbf{g'} \right)   \sum_{\mathbf{g''}} \frac{\partial B_{\mu,\nu,\tau,\omega}^\mathbf{0,g,g'',g''+g'}  }{\partial \mathbf{A}_\eta} \Bigg] \nonumber \\
&+& \frac{\partial \left[ P_{\mu \nu}^{\sigma \sigma^\prime} \left( \mathbf{g} \right) \right]^\ast }{\partial \mathbf{A}_\eta} H_{\mu \nu}^{\sigma \sigma^\prime} \left( \mathbf{g} \right) \Bigg\}  \; .
\end{eqnarray}
The explicit calculation of the derivative of the density matrix in the last line of Eq. (\ref{eqn:gradE}) can be avoided by first making use of Eqs. (\ref{eqn:pmat}) and (\ref{eqn:fourier}), leading to:
\begin{widetext}
\begin{eqnarray}
\label{eqn:dP_dev}
\sum_\mathbf{g} \sum_{\mu \nu} \frac{\partial \left[ P_{\mu \nu}^{\sigma \sigma^\prime} \left( \mathbf{g} \right) \right]^\ast  }{\partial \mathbf{A}_\eta} H_{\mu \nu}^{\sigma \sigma^\prime} \left( \mathbf{g} \right) &=& \sum_{\mu \nu} \frac{1}{\Omega} \sum_i^\text{bands} f_i \int_\Omega d \mathbf{k} \ \theta\left[ \varepsilon_F - \epsilon_i ({\mathbf{k}}) \right] \left\{ \frac{\partial \left[ C^{\sigma}_{\mu,i} \left( \mathbf{k} \right) \right]^\ast}{\partial \mathbf{A}_\eta}  H_{\mu \nu}^{\sigma \sigma^\prime} \left( \mathbf{k} \right)   C^{\sigma^\prime}_{\nu,i} \left( \mathbf{k} \right) + c.c. \right\} \nonumber \\
&=& \sum_{\mu \nu} \frac{1}{\Omega} \sum_i^\text{bands} f_i \int_\Omega d \mathbf{k} \ \theta\left[ \varepsilon_F - \epsilon_i ({\mathbf{k}}) \right] \left\{ \frac{\partial \left[ C^{\sigma}_{\mu,i} \left( \mathbf{k} \right) \right]^\ast}{\partial \mathbf{A}_\eta}  \delta_{\sigma, \sigma^\prime} \left( \epsilon_i \left( \mathbf{k} \right) + Q \right) S_{\mu \nu}^{\sigma \sigma} \left( \mathbf{k} \right)    C^{\sigma^\prime}_{\nu,i} \left( \mathbf{k} \right) + c.c. \right\} \; ,
\end{eqnarray}
\end{widetext}
where we have made use of Eq. (\ref{eqn:GKS_mod}). Furthermore, a differentiation of Eq. (\ref{eqn:ortho}) provides:
\begin{eqnarray}
\label{eqn:ortho_diff}
0&=& \delta_{\sigma, \sigma^\prime}  \sum_{\mu \nu} \frac{\partial \left[ C^{\sigma}_{\mu,i} \left( \mathbf{k} \right) \right]^\ast}{\partial \mathbf{A}_\eta} S_{\mu \nu}^{\sigma \sigma} \left( \mathbf{k} \right) C^{\sigma^\prime}_{\nu,i} \left( \mathbf{k} \right) \nonumber \\
&+& C^{\sigma}_{\mu,i} \left( \mathbf{k} \right) \frac{ \partial S_{\mu \nu}^{\sigma \sigma} \left( \mathbf{k} \right)}{\partial \mathbf{A}_\eta} C^{\sigma^\prime}_{\nu,i} \left( \mathbf{k} \right) + C^{\sigma}_{\mu,i} \left( \mathbf{k} \right) S_{\mu \nu}^{\sigma \sigma} \left( \mathbf{k} \right) \frac{\partial C^{\sigma^\prime}_{\nu,i} \left( \mathbf{k} \right)}{\partial \mathbf{A}_\eta} \; .
\end{eqnarray}
Inserting Eqs. (\ref{eqn:dP_dev}) and (\ref{eqn:ortho_diff}) into Eq. (\ref{eqn:gradE}), we obtain:
\begin{eqnarray}
\label{eqn:gradE2}
\frac{\partial E}{\partial \mathbf{A}_\eta} = \frac{1}{2} \sum_\mathbf{g} \sum_{\sigma,\sigma^\prime} \sum_{\mu \nu} \Bigg\{  \left[ P_{\mu \nu}^{\sigma \sigma^\prime} \left( \mathbf{g} \right) \right]^\ast \Bigg[ 2 \frac{\partial  h^{\sigma \sigma^\prime}_{\mu \nu} \left( \mathbf{g} \right)  }{\partial \mathbf{A}_\eta} \nonumber \\
+ \frac{\partial  V^{\sigma \sigma^\prime}_{\mu \nu} \left( \mathbf{g} \right)  }{\partial \mathbf{A}_\eta}    +  \sum_{\tau \omega} \sum_{\sigma^{\prime \prime},\sigma^{\prime \prime \prime}}  \sum_\mathbf{g'} P_{\tau \omega}^{\sigma^{\prime \prime} \sigma^{\prime \prime \prime}} \left( \mathbf{g'} \right)   \sum_{\mathbf{g''}} \frac{\partial B_{\mu,\nu,\tau,\omega}^\mathbf{0,g,g'',g''+g'}  }{\partial \mathbf{A}_\eta} \Bigg] \nonumber \\
- \left[ W_{\mu \nu}^{\sigma \sigma} \left( \mathbf{g} \right) \right]^\ast \frac{\partial  S_{\mu \nu}^{\sigma \sigma} \left( \mathbf{g} \right) }{\partial \mathbf{A}_\eta} \Bigg\} \; ,
\end{eqnarray}
in which $W_{\mu \nu}^{\sigma \sigma} \left( \mathbf{g} \right)$ are elements of the direct-space energy-weighted density-matrix:
\begin{eqnarray}
\label{eqn:Wmat}
W_{\mu \nu}^{\sigma \sigma^\prime} \left( \mathbf{g} \right) &=& \frac{1}{\Omega} \sum_i^\text{bands} f_i  \int_\Omega d \mathbf{k} \ \left( \epsilon_i ({\mathbf{k}})  + Q \right) e^{\i \mathbf{k} \cdot \mathbf{g}} \nonumber \\
&\times& C^{\sigma}_{\mu,i} \left( \mathbf{k} \right) \left[ C^{\sigma^\prime}_{\nu,i} \left( \mathbf{k} \right) \right]^\ast \theta\left[ \varepsilon_F - \epsilon_i ({\mathbf{k}}) \right]  \; .
\end{eqnarray}

\subsection{Analytical Gradients with Respect to Lattice Vectors}

Having determined derivatives with respect to atomic displacements $\mathbf{A}_\eta$, the treatment may be extended to derivatives with respect to direct lattice vectors $\mathbf{a}_1$, $\mathbf{a}_2$ and $\mathbf{a}_3$. This may be achieved, by first writing a general position in the lattice basis:
\begin{equation}
\label{eqn:frac_coord}
\mathbf{A}_\eta + \mathbf{g} = \sum_{i=1}^3 \left( f_{\eta,i} + n_{\mathbf{g},i} \right) \mathbf{a}_i
\end{equation}
in which $f_{\eta,1}, f_{\eta,2}, f_{\eta,3} \in \mathbb{R}$ and $n_{\mathbf{g},1}, n_{\mathbf{g},2}, n_{\mathbf{g},3} \in \mathbb{Z}$ are, respectively real and integer quantities. Starting from Eq. (\ref{eqn:E}), and proceeding as in Section \ref{sec:grad_atom} yields:
\begin{eqnarray}
\label{eqn:gradcell}
\frac{\partial E}{\partial \mathbf{a}_i} = \frac{1}{2} \sum_\mathbf{g} \sum_{\sigma,\sigma^\prime} \sum_{\mu \nu} \Bigg\{  \left[ P_{\mu \nu}^{\sigma \sigma^\prime} \left( \mathbf{g} \right) \right]^\ast \Bigg[ 2 \frac{\partial  h^{\sigma \sigma^\prime}_{\mu \nu} \left( \mathbf{g} \right)  }{\partial \mathbf{a}_i} \nonumber \\
+ \frac{\partial  V^{\sigma \sigma^\prime}_{\mu \nu} \left( \mathbf{g} \right)  }{\partial \mathbf{a}_i}    +  \sum_{\tau \omega} \sum_{\sigma^{\prime \prime},\sigma^{\prime \prime \prime}}  \sum_\mathbf{g'} P_{\tau \omega}^{\sigma^{\prime \prime} \sigma^{\prime \prime \prime}} \left( \mathbf{g'} \right)   \sum_{\mathbf{g''}} \frac{\partial B_{\mu,\nu,\tau,\omega}^\mathbf{0,g,g'',g''+g'}  }{\partial \mathbf{a}_i} \Bigg] \nonumber \\
- \left[ W_{\mu \nu}^{\sigma \sigma} \left( \mathbf{g} \right) \right]^\ast \frac{\partial  S_{\mu \nu}^{\sigma \sigma} \left( \mathbf{g} \right) }{\partial \mathbf{a}_i} \Bigg\} \; .
\end{eqnarray}

\subsection{Derivatives of SOC Integrals}

In Eqs. (\ref{eqn:gradE}) and (\ref{eqn:gradcell}), we require analytical derivatives of the integrals with respect to atomic displacements and lattice vectors. Their calculation (apart from SOC integrals) has been described elsewhere.\cite{desmarais2023efficient,GORBS,Doll1,Doll2,DOLL_CELLGRAD1,DOLL_CELLGRAD2,DOLL_STRESS} In the following sections we concentrate on the calculation of analytical derivatives of SOC integrals.

\subsubsection{Derivatives of SOC Integrals: Atomic Displacements}

As reported in Ref. \onlinecite{desmarais2019spin}, the explicit energy contribution in Eq. (\ref{eqn:E}) from the SOC operator of Eq. (\ref{eqn:fock_1e}) can be written in the following computationally convenient way:
\begin{eqnarray}
\label{eqn:e_soc}
E_{SO} &=& - 2 \Re \sum_{\mathbf{g} } \sum_{\mu \ge \nu} \Big\{ u_{SO,\mu \nu}^{\alpha \alpha} \left(  \mathbf{g} \right)  \left[ P_{\mu \nu}^{\alpha \alpha} \left(  \mathbf{g} \right)  -  P_{\mu \nu}^{\beta \beta} \left(  \mathbf{g} \right)    \right]^\ast \nonumber \\
&-& u_{SO,\mu \nu}^{\alpha \beta} \left(  \mathbf{g} \right)  \left[ P_{\mu \nu}^{\alpha \beta} \left(  \mathbf{g} \right)  -  P_{\mu \nu}^{\beta \alpha} \left(  \mathbf{g} \right)    \right]^\ast   \Big\} \; ,
\end{eqnarray}
in which the diagonal spin-blocks of the SOC matrix elements read (in the given RECP approximation):\cite{desmarais2019spin}
\begin{eqnarray}
\label{eqn:uSO_diag}
u_{SO,\mu \nu}^{\alpha \alpha} \left(  \mathbf{g} \right) &=& \sum_\mathbf{g''} \sum_c^\text{atoms} \sum_{l=0}^\mathcal{L} \langle \chi_{\mu,\mathbf{0}} \vert  \hat{\xi}_{l,c,\mathbf{g''}} \nonumber \\
&\times& \hat{P}_{l,c,\mathbf{g''}}  \hat{L}_{z,l,c,\mathbf{g''}}  \hat{P}_{l,c,\mathbf{g''}}   \vert \chi_{\nu,\mathbf{g}}  \rangle \nonumber \\
&\equiv& \sum_\mathbf{g''} \sum_c^\text{atoms} \sum_{l=0}^\mathcal{L} \langle \chi_{\mu,\mathbf{0}} \vert  \hat{u}_{SO,c,\mathbf{g''}}^{\alpha \alpha}    \vert \chi_{\nu,\mathbf{g}}  \rangle \nonumber \\
&\equiv& \sum_\mathbf{g''} \sum_c^\text{atoms} u_{SO,\mu \nu}^{\alpha \alpha} \left(  \mathbf{g};c,\mathbf{g''} \right) \; ,
\end{eqnarray}
where $\hat{\xi}_{l,c,\mathbf{g''}}$ are radial operators centered at $\mathbf{A}_c$ in cell $\mathbf{g''}$, $\hat{P}_{l,c,\mathbf{g''}}$ are projectors onto real spherical harmonics and $\hat{L}_{z,l,c,\mathbf{g''}}$ is the (pure-imaginary) $z$-component electron-nuclear angular-momentum operator. In our implementation, $\mathcal{L}$ has a maximum value of 4 ($g$-type projectors). Off-diagonal spin-blocks are written:\cite{desmarais2019spin}
\begin{eqnarray}
\label{eqn:uSO_offdiag}
u_{SO,\mu \nu}^{\alpha \beta} \left(  \mathbf{g} \right) &=& \sum_\mathbf{g''} \sum_c^\text{atoms} \sum_{l=0}^\mathcal{L} \langle \chi_{\mu,\mathbf{0}} \vert  \hat{\xi}_{l,c,\mathbf{g''}} \hat{P}_{l,c,\mathbf{g''}}  \nonumber \\
&\times& \hat{L}_{-,l,c,\mathbf{g''}}  \hat{P}_{l,c,\mathbf{g''}}   \vert \chi_{\nu,\mathbf{g}}  \rangle \nonumber \\
&\equiv& \sum_\mathbf{g''} \sum_c^\text{atoms} \langle \chi_{\mu,\mathbf{0}} \vert  \hat{u}_{SO,c,\mathbf{g''}}^{\alpha \beta}   \vert \chi_{\nu,\mathbf{g}}  \rangle \nonumber \\
&\equiv& \sum_\mathbf{g''} \sum_c^\text{atoms} u_{SO, \mu \nu}^{\alpha \beta} \left(  \mathbf{g};c,\mathbf{g''} \right) \; ,
\end{eqnarray}
where $\hat{L}_{-,l,c,\mathbf{g''}}$ is the corresponding angular-momentum annihilation operator. Comparing Eq. (\ref{eqn:e_soc}) with Eq. (\ref{eqn:gradE}), the terms associated with derivatives of the SOC integrals are then:
\begin{eqnarray}
\label{eqn:e_socprime}
E_{SO}^\prime &=& -2 \Re \sum_{\mathbf{g} } \sum_{\mu \ge \nu} \Big\{ \frac{ \partial u_{SO,\mu \nu}^{\alpha \alpha} \left(  \mathbf{g} \right)}{\partial \mathbf{A}_\eta}  \left[ P_{\mu \nu}^{\alpha \alpha} \left(  \mathbf{g} \right)  -  P_{\mu \nu}^{\beta \beta} \left(  \mathbf{g} \right)    \right]^\ast \nonumber \\
&-& \frac{\partial u_{SO,\mu \nu}^{\alpha \beta} \left(  \mathbf{g} \right)}{ \partial \mathbf{A}_\eta  }  \left[ P_{\mu \nu}^{\alpha \beta} \left(  \mathbf{g} \right)  -  P_{\mu \nu}^{\beta \alpha} \left(  \mathbf{g} \right)    \right]^\ast   \Big\} \; .
\end{eqnarray}
To calculate the necessary derivative integrals, we employ integration by parts to develop the following translational invariance sum rule, involving derivatives with respect to the centers of the bra- $\mathbf{A}_\mu$, ket- $\mathbf{A}_\nu$ and operator $\mathbf{A}_c$:
\begin{equation}
\label{eqn:trans_inv}
\frac{\partial}{ \partial \mathbf{A}_\mu} + \frac{\partial}{ \partial \mathbf{A}_\nu} + \frac{\partial}{ \partial \mathbf{A}_c} =0 \; .
\end{equation}
Inserting Eq. (\ref{eqn:trans_inv}) into Eqs. (\ref{eqn:uSO_diag}) and (\ref{eqn:uSO_offdiag}), we obtain:
\begin{eqnarray}
\label{eqn:uSO_deriv}
\frac{ \partial u_{SO,\mu \nu}^{\sigma \sigma^\prime} \left(  \mathbf{g};c,\mathbf{g''} \right)}{\partial \mathbf{A}_\eta} = \langle \frac{\partial \chi_{\mu,\mathbf{0}}}{\partial \mathbf{A}_\mu  } \vert  \hat{u}_{SO, c,\mathbf{g''}}^{\sigma \sigma^\prime} \vert \chi_{\nu,\mathbf{g}}  \rangle \left( \delta_{\eta,\mu} - \delta_{\eta,c} \right) \nonumber \\
+\langle \chi_{\mu,\mathbf{0}} \vert  \hat{u}_{SO,c,\mathbf{g''}}^{\sigma \sigma^\prime} \vert \frac{\partial \chi_{\nu,\mathbf{g}}}{\partial \mathbf{A}_\nu  }  \rangle \left( \delta_{\eta,\nu} - \delta_{\eta,c} \right) \; .
\end{eqnarray}

\subsubsection{Derivatives of SOC Integrals: Lattice Vectors}

Insertion of Eq. (\ref{eqn:frac_coord}) into Eqs. (\ref{eqn:uSO_diag}) and (\ref{eqn:uSO_offdiag}) permits to write derivatives of the SOC integrals with respect to lattice vectors required in Eq. (\ref{eqn:gradcell}):
\begin{widetext}
\begin{eqnarray}
\label{eqn:uSO_deriv_cell}
\frac{ \partial u_{SO,\mu \nu}^{\sigma \sigma^\prime} \left(  \mathbf{g};c,\mathbf{g''} \right)}{\partial \mathbf{a}_{i}} &=& \frac{ \partial u_{SO,\mu \nu}^{\sigma \sigma^\prime} \left(  \mathbf{g};c,\mathbf{g''} \right)}{\partial \mathbf{A}_{\mu}} \frac{\partial \mathbf{A}_{\mu}  }{\partial \mathbf{a}_{i}  } 
+ \frac{ \partial u_{SO,\mu \nu}^{\sigma \sigma^\prime} \left(  \mathbf{g};c,\mathbf{g''} \right)}{\partial \left( \mathbf{A}_{\nu} + \mathbf{g} \right) } \frac{\partial \left( \mathbf{A}_{\nu} + \mathbf{g} \right)  }{\partial \mathbf{a}_{i}  } 
+ \frac{ \partial u_{SO,\mu \nu}^{\sigma \sigma^\prime} \left(  \mathbf{g};c,\mathbf{g''} \right)}{\partial \left( \mathbf{A}_{c} + \mathbf{g''} \right) } \frac{\partial \left( \mathbf{A}_{c} + \mathbf{g''} \right)  }{\partial \mathbf{a}_{i}  } \nonumber \\
&=&  \langle \frac{\partial \chi_{\mu,\mathbf{0}}}{\partial \mathbf{A}_{\mu}  } \vert  \hat{u}_{SO, c,\mathbf{g''}}^{\sigma \sigma^\prime} \vert \chi_{\nu,\mathbf{g}}  \rangle \left( f_{\mu,i} - f_{c,i}  -  n_{\mathbf{g''},i} \right) 
+\langle \chi_{\mu,\mathbf{0}} \vert  \hat{u}_{SO,c,\mathbf{g''}}^{\sigma \sigma^\prime} \vert \frac{\partial \chi_{\nu,\mathbf{g}}}{\partial \mathbf{A}_{\nu}  }  \rangle \left( f_{\nu,i} + n_{\mathbf{g},i} - f_{c,i}  -  n_{\mathbf{g''},i} \right) \; ,
\end{eqnarray}
\end{widetext}
where use has been made of Eq. (\ref{eqn:trans_inv}).

\section{Computational Details}
\label{sec:comput}

All calculations are performed with a developmental version of the \textsc{Crystal} program.\cite{CRYSTAL17PAP,erba2022crystal23} The SVWN5 exchange-correlation (xc) functional of the LDA, PBE xc functional of the GGA, and PBE0 xc functional of the global hybrid GGA are used.\cite{dirac1930note,vosko1980accurate,pbe_art,pbe0_art} Large-core pseudo-potentials are used for Po, and both large- and small-core pseudo-potentials for I.\cite{cao2011pseudopotentials} For the large-core calculations, valence basis sets for I and Po of the form (6$s$5$p$2$d$)/[4$s$3$p$2$d$] and (4$s$4$p$)/[2$s$2$p$] have been modified from the ones originally presented in Ref. \onlinecite{stoll2002relativistic}, respectively. For the molecular small-core calculations, the triple-zeta valence basis set for I of Ref. \onlinecite{peterson2006spectroscopic} is used. The basis set for H is taken from Ref. \onlinecite{gatti1994crystal}. For the periodic calculations, we use small-core pseudo-potentials and basis sets modified from Ref. \onlinecite{laun2022bsse}. For application to the I$_2$ and  CsI$_3$ crystals, reciprocal space is sampled in a $10 \times 10 \times 10$ and $4 \times 4 \times 4$ Monkhorst-Pack net, respectively. A tolerance of 10$^{-8}$ Hartree on the total energy is used as a convergence criterion for the self-consistent field (SCF) procedure. The five \texttt{TOLINTEG} parameters that control truncation of the Coulomb and exact-exchange infinite series are set to 8 8 8 8 20. The xc functional and potential (in their collinear spin-DFT formulation) are sampled on a direct-space pruned grid over the unit-cell volume with Lebedev angular and Gauss-Legendre radial quadratures, employing 99 radial and 1454 angular points (keyword \texttt{XXLGRID}). Both the atomic fractional coordinates and lattice parameters are fully optimized with analytical gradients of the total energy and a quasi-Newton scheme in the Broyden-Fletcher-Goldfarb-Shanno (BFGS) variant.\cite{Doll1,Doll2,DOLL_CELLGRAD2,CivaGRAD} The initial guess for the Hessian of the BFGS scheme is taken as the identity matrix. Full input decks are available in \textsc{Crystal} format in the ESI.\cite{supp_mat}

\begin{table}[t!]
\footnotesize
\caption{Comparison between analytical and numerical cell gradient for the infinite 1D chain of H$_2$Po$_2$ in the presence of SOC, for EXX and LDA calculations. The analytical cell gradient $G_a$ is reported along with its SOC contribution $G_a^\textup{SOC}$. Differences $\Delta$ are reported between the analytical and numerical gradient as obtained through different finite difference schemes: a two-point, one-sided formula (2O), a two-point, two-sided formula (2T), and a four-point, two-sided formula (4T). All values in Hartree/bohr.}
\label{tab:h2po2_numana}
\vspace{2pt}
\begin{tabular}{lccccc}
\hline
\hline
& $G_a$ & $G_a^\textup{SOC}$ & $\Delta_\text{2O}$ & $\Delta_\text{2T}$  & $\Delta_\text{4T}$  \\
&&\\
HF  & 5.86$\times 10^{-2}$ & 5.26$\times 10^{-3}$  & 2.26$\times 10^{-4}$ &  5.00$\times 10^{-6}$ &  4.95$\times 10^{-6}$ \\ 
LDA & 5.83$\times 10^{-2}$ & 7.63$\times 10^{-3}$  & 2.27$\times 10^{-4}$ &   1.30$\times 10^{-5}$ & 1.30$\times 10^{-5}$\\
&&\\
\hline
\hline
\end{tabular}
\end{table}

\section{Results and Discussion}

We discuss the numerical accuracy of the analytical forces relative to numerical ones on a simple model system: a periodic 1D chain of H$_2$Po$_2$ units. We discuss the effect of SOC on structural parameters of the I$_2$ molecule, I$_2$ orthorhombic molecular crystal, and Caesium triiodide orthorhombic crystal. The effect of renormalization of the electron-electron interaction through SOC-induced spin-currents is quantified.

\subsection{Numerical Validation on a Model System}

To validate our approach for computation of analytical gradients of the total energy in SCDFT, and demonstrate its high numerical accuracy, we perform calculations on a model system represented by the infinite 1D chain of H$_2$Po$_2$. This system is chosen based on the very large contribution of SOC to the forces. We compare the analytical cell gradient against numerical computations from finite differences of the total energy. The employed geometry for the H$_2$Po$_2$ chain, as well as the values of the finite difference parameters are provided in the ESI. We present such comparison in Table \ref{tab:h2po2_numana} for two types of GKS calculations: exact exchange approximation (EXX) and LDA. The computed analytical cell gradient $G_a$ is 5.861 $\times 10^{-2}$ for EXX and 5.826 $\times 10^{-2}$ Hartree/bohr for LDA. The total SOC contribution to the cell gradient $G_a^\textup{SOC}$ amounts to 5.265 $\times 10^{-3}$ and 7.628 $\times 10^{-3}$, respectively. Thus, the effect of SOC on the cell gradient is of the order of 10-15\%. 

Different numerical finite difference schemes are used to compare the analytical forces with: a two-point, one-sided formula (2O), a two-point, two-sided formula (2T), and a four-point, two-sided formula (4T). Differences between the analytical force and the numerical one are already on the order of $10^{-4}$ a.u. when the simplest 2O formula is used (i.e. being around one order of magnitude smaller than the SOC contribution to the gradient). These differences are further reduced to $10^{-5}$ a.u. (in the case of LDA) and even $10^{-6}$ a.u. (in the case of EXX) when more robust 2T or 4T numerical schemes are used, thus demonstrating the high numerical accuracy of our analytical approach to energy gradients within the SCDFT. A better agreement is obtained in the case of EXX, wherein the implementation is fully analytical (aside from integration over the first Brillouin zone, as well as diagonalization of the secular GKS equation). In contrast, the LDA computation contains an additional numerical step: the integration of the xc energy-density and potential over the direct-space unit-cell volume. In this case, imperfect cancellation of errors in the numerical quadrature slightly worsens the agreement.

\begin{figure}[t!]
\centering
\includegraphics[width=8.0cm]{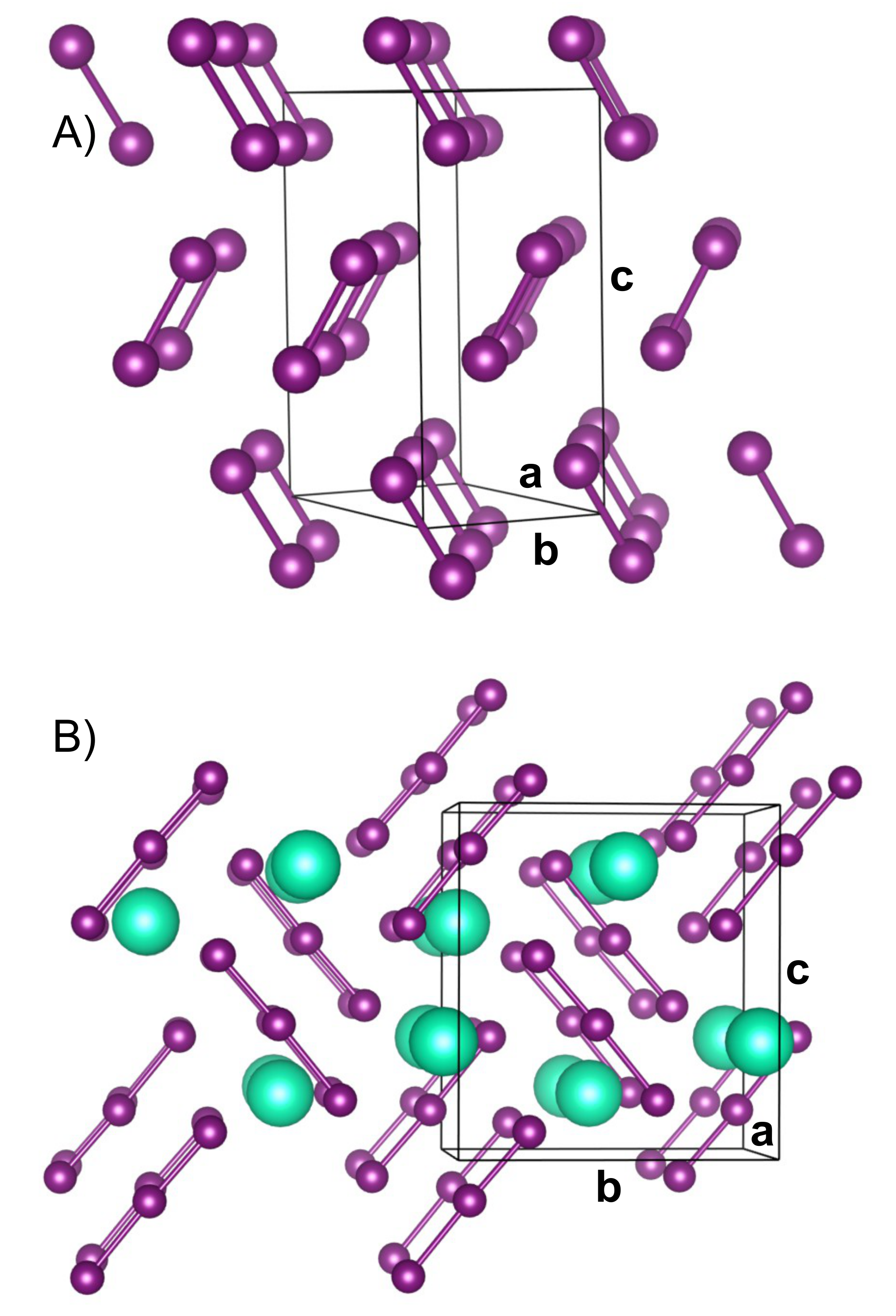}
\caption{Crystal structure of A) the I$_2$ orthorhombic molecular crystal, and B) the CsI$_3$ orthorhombic crystal.}
\label{fig:crystals}
\end{figure}

\subsection{Application to the I$_2$ Molecule}

We discuss the effect of SOC on the bond length of the diiodide molecule by comparing our SCDFT approach to simpler (S)DFT ones where the effect of the spin-current densities on the renormalization of the electron-electron interaction is omitted. We study the molecule both in its neutral ground state I$_2$ (closed-shell configuration, i.e. time-reversal symmetry preserving) and in its anionic form I$_2^-$ (open-shell configuration, i.e. time-reversal symmetry breaking). We perform calculations using both the PBE and PBE0 xc functionals, and both large-core (LC) and small-core (SC) pseudo-potentials. Results are reported in Table \ref{tab:I2_mol}. The experimental gas phase bond length at -80 $^\circ$C, for comparison, is around 2.674 \AA{}.\cite{buontempo1997determination,ukaji1966effect} The best agreement with the experimental figure is obtained from the PBE0 SC calculation, which gives 2.678 \AA{} with our SCDFT approach. 

Inspection of the Table clearly shows that SOC systematically induces the lengthening of the bond in all cases. For the I$_2$ species, an SCDFT treatment of SOC (i.e. including spin-current densities) results in a bond lengthening that is twice as large as that from a DFT approach where spin-current densities are neglected. This is observed consistently in both LC and SC calculations. Thus, around half of the effect of SOC on the ground state geometry of the I$_2$ molecule is accounted for by modification of the electron-electron interaction through SOC-induced spin-current densities.

\begin{table}[t!]
\caption{Equilibrium bond length of the I$_2$ and I$_2^-$ molecules, as computed with the PBE and PBE0 xc functionals, with both large-core (LC) and small-core (SC) pseudo-potentials. The scalar relativistic (SR) value is reported (i.e. obtained before inclusion of SOC) along with the effect of SOC ($\Delta_\text{SOC}$). All values are in \AA. The experimental gas phase bond length of I$_2$ at -80 $^\circ$C, for comparison, is around 2.674 \AA.\cite{buontempo1997determination,ukaji1966effect}}
\vspace{2pt}
\label{tab:I2_mol}
\begin{tabular}{lccccccc}
\hline
\hline
& \multicolumn{3}{c}{I$_2$} && \multicolumn{3}{c}{I$_2^-$} \\
\cline{2-4}\cline{6-8}
& SR & \multicolumn{2}{c}{$ \Delta_\text{SOC} $} && SR & \multicolumn{2}{c}{$ \Delta_\text{SOC} $}\\
\cline{3-4}\cline{7-8}
& DFT & DFT  & SCDFT  && SDFT & SDFT  & SCDFT \\
&&\\
PBE  (LC) & 2.822 & 0.034 & - && 3.399 & 0.064 & - \\
PBE0 (LC) & 2.792 & 0.011 & 0.026 && 3.339 & 0.047 & 0.024 \\
&&\\
PBE  (SC) & 2.694 & 0.019 & - && 3.315 & 0.049 & - \\
PBE0 (SC) & 2.663 & 0.008 & 0.015 && 3.252 & 0.036 &  0.018 \\
&&\\
\hline
\hline
\end{tabular}
\end{table}

The last three columns of Table \ref{tab:I2_mol} provide data for the I$_2^-$ species. As anticipated before, also in this case SOC produces the lengthening of the bond, but with an important difference with respect to the case of I$_2$ when it comes to the role played by the spin-current densities. Now, by neglecting the effect of the SOC-induced spin-current densities the effect of SOC on the bond length would be largely overestimated by the SDFT, with a lengthening by 0.047 and 0.036 \AA{} with PBE0 with LC and SC, respectively, which is reduced to 0.024 and 0.018 \AA{} by inclusion of the spin-current densities within the SCDFT. The effect of SOC on bond lengthening is thus overestimated by 100 \% if spin-current densities are not included in the xc functional.

\begin{table}[b!]
\caption{Structural parameters of the I$_2$ orthorhombic molecular crystal computed with the PBE0 xc functional and SC pseudo-potentials. Absolute values are reported from the scalar relativistic (SR) calculation. The effect of SOC, $\Delta_\text{SOC}$, is reported for both a DFT and SCDFT treatment. Data are reported in the primitive basis of the lattice. Low-temperature (5 K) experimental values are taken from Ref. \onlinecite{ibberson1992high}.}
\vspace{5pt}
\label{I2_cry}
\begin{tabular}{lcccc}
\hline
\hline
                               & Exp. &  SR &  \multicolumn{2}{c}{$\Delta_\text{SOC}$}  \\
                               \cline{4-5}
                                   & &   & DFT & SCDFT  \\
                                   &&\\
a/{\AA}                   & 4.254    &  4.316   & -0.036  & -0.033  \\     
c/{\AA}                    & 9.796    &   9.711   & 0.035  &0.041 \\
c/a                           & 2.302    &   2.250   & 0.0270  & 0.0272 \\
$\gamma$/{$^\circ$}& 113.584 &   115.550 & 0.104  &  -0.004 \\
V/{\AA$^3$}            & 162.482  & 163.223 & -2.241  & -1.857 \\
x/a                         & 0.1549    &   0.1589  & 0.002 & 0.002 \\
z/c                         & 0.1175    &   0.1186  & 0.000 & 0.000 \\
I-I/{\AA}                &  2.717    &   2.728   &  0.011 & 0.018 \\ 
&&\\
\hline
\hline
\end{tabular}
\end{table}

\subsection{Application to the I$_2$ Molecular Crystal}

We now discuss the application of our approach to the I$_2$ orthorhombic molecular crystal. The structure of the crystal (in terms of its conventional lattice cell) is depicted in Fig. \ref{fig:crystals} A). Low temperature (5 K) experimental structural data are available \cite{ibberson1992high} and are reported in Table \ref{I2_cry}, along with our optimized theoretical structural parameters (in terms of the primitive lattice cell). The hybrid PBE0 xc functional is used here both in a DFT and SCDFT framework. Absolute values are reported from the scalar relativistic (SR) calculation. The effect of SOC, $\Delta_\text{SOC}$, is highlighted in the last two columns.

Based on the experiments, the I-I bond length in the crystal, 2.717 \AA{}, is larger than in gas phase, 2.674 \AA{}. Our theoretical calculations are consistent with this picture. Also in the crystal, SOC induces the lengthening of the bond with an elongation that is nearly twice as large in SCDFT (0.018 \AA{}) than it is in DFT (0.011 \AA{}). Compared to the gas phase calculations, we observe that the SOC-induced bond elongation is increased by around 17\%. The $\mathbf{a}$ and $\mathbf{b}$ lattice parameters (in the plane perpendicular to the nearest-neighbour I-I bond) are shortened by SOC, while the $\mathbf{c}$ lattice parameter is lengthened by SOC. An overall SOC-induced volume contraction by -2.241 {\AA$^3$} in DFT decreased, in absolute value, to -1.857  {\AA$^3$} in SCDFT (corresponding to a volume expansion of about 1.1\% with respect to the SR calculation).

\begin{table}[t!]
\caption{Structural parameters of the CsI$_3$ orthorhombic crystal (space group P$mcn$) computed with the PBE0 xc functional. Absolute values are reported from the scalar relativistic (SR) calculation. The effect of SOC, $\Delta_\text{SOC}$, is reported for both a DFT and SCDFT treatment. Data are reported in the conventional basis of the lattice. Low-temperature (113 K) experimental values are taken from Ref. \onlinecite{runsink1972refinement}.}
\label{CsI3_struct}
\vspace{5pt}
\begin{tabular}{lcccc}
\hline
\hline
                               & Exp. &  SR &  \multicolumn{2}{c}{$\Delta_\text{SOC}$}  \\
                               \cline{4-5}
                                   & &   & DFT & SCDFT  \\
                                   &&\\
a/{\AA}             & 6.751  &  6.920 & -0.020  & -0.033\\
b/{\AA}             & 9.963  & 10.171 &  0.000  & -0.008\\
c/{\AA}             & 10.997 & 10.961 &  0.005 & -0.001\\
V/{\AA$^3$}     & 739.661& 771.544& -2.020  & -4.357\\
Cs-I/{\AA}         & 3.709 &  3.796 &  0.000 & -0.003\\
I(1)-I(2)/{\AA}    & 2.842 &  2.860 &  0.011 & 0.013\\ 
I(2)-I(3)/{\AA}    & 3.038 &  2.998 &  0.015  & 0.012\\
&&\\
\hline
\hline
\end{tabular}
\end{table}

\subsection{Application to the CsI$_3$ Crystal}

The structure of the CsI$_3$ orthorhombic (space group P$mcn$) crystal is depicted in Fig. \ref{fig:crystals} B) with four Cs and twelve I atoms in the unit cell. The crystal exhibits linear I$_3$ molecules characterized by a small asymmetry in terms of the two I-I bond lengths, with a shorter one of 2.842 \AA{} and a longer one of 3.038 \AA{} at -160 $^\circ$C.\cite{runsink1972refinement} The asymmetry may be effectively tuned by external stimuli, such as temperature or pressure.\cite{poreba2022investigating} Low temperature (113 K) experimental structural data are reported in Table \ref{CsI3_struct}, along with our optimized theoretical structural parameters. The hybrid PBE0 xc functional is used here both in a DFT and SCDFT framework. Absolute values are reported from the scalar relativistic (SR) calculation. The effect of SOC, $\Delta_\text{SOC}$, is highlighted in the last two columns.

The effect of SOC on the structure of this crystal is more articulated than it was on the I$_2$ molecular crystal. Indeed, while SOC still induces the elongation of I-I bonds, it shortens Cs-I interactions, particularly so within an SCDFT description including SOC-induced spin-current densities. This results in an overall volume contraction by 2.020 {\AA$^3$} from DFT and 4.357  {\AA$^3$}  from SCDFT (corresponding to a volume contraction of about 0.6\% with respect to the SR calculation). SOC is also found to contract the structure anisotropically, with the largest contraction occurring along the {\bf a} lattice parameter. This is consistent with 
{\bf a} being a crystallographic direction with no I-I bond components: indeed, I-I bonds are rather oriented in the {\bf bc} plane, as shown in Fig. \ref{fig:crystals} B).

\section{Conclusion}

We have presented analytical gradients of the total energy for local-density (LDA) and generalized-gradient (GGA) hybrid approximations to generalized Kohn-Sham spin-current density functional theory (GKS-SCDFT), including spin-orbit coupling (SOC). Our strategy has been implemented in a developmental version of the \textsc{Crystal} program. The numerical accuracy of the analytical forces has been validated against forces obtained by different finite difference schemes. Application on the I$_2$ and CsI$_3$ crystals has shown that terms in the exchange-correlation functional arising from SOC-induced spin-current densities, as accounted for within an SCDFT framework, lead to significant structural changes. These are reflected on a lattice expansion or contraction which account (in the case of the CsI$_3$ crystal) to more than half of the total effect due to SOC. Future efforts will be devoted to inclusion of explicit contributions in density functional approximations from modification of the curvature of the exchange-correlation hole by SOC-induced current densities, which must be taken into account at the level of meta-GGA approximations to GKS-SCDFT.

\acknowledgments

This research has received funding from the Project CH4.0 under the MUR program ``Dipartimenti di Eccellenza 2023-2027'' (CUP: D13C22003520001).


\providecommand{\latin}[1]{#1}
\makeatletter
\providecommand{\doi}
  {\begingroup\let\do\@makeother\dospecials
  \catcode`\{=1 \catcode`\}=2 \doi@aux}
\providecommand{\doi@aux}[1]{\endgroup\texttt{#1}}
\makeatother
\providecommand*\mcitethebibliography{\thebibliography}
\csname @ifundefined\endcsname{endmcitethebibliography}
  {\let\endmcitethebibliography\endthebibliography}{}

\end{document}